\documentclass[a4paper,11pt]{article}
\usepackage{cite}
\usepackage{fullpage,setspace,hyperref,comment,caption}
\usepackage[font=footnotesize,labelfont=bf,margin=1cm]{caption}
\usepackage{cite} 
\usepackage{color}
\usepackage{amsfonts}
\usepackage{amsmath}
\usepackage{amssymb}
\usepackage{todonotes} 
\usepackage{tikz}
\usepackage{tikz-cd}
\usepackage[utf8]{inputenc}
\usepackage{slashed}
\usepackage{tcolorbox}
\usepackage{bbm}
 \usepackage[normalem]{ulem}

\newcommand{\beq}{\begin{equation}}
\newcommand{\eeq}{\end{equation}}
\newcommand{\be}{\begin{equation}}
\newcommand{\ee}{\end{equation}}

\newcommand{\eda}{\mathfrak{d}_{E}}

\newcommand{\cA}{{\cal A}}
\newcommand{\cB}{{\cal B}}
\newcommand{\cC}{{\cal C}}
\newcommand{\CD}{{\cal D}}
\newcommand{\cE}{{\cal E}}
\newcommand{\cF}{{\cal F}}

\newcommand{\obf}[1]{\overline{\mathbf{#1}}}
\newcommand{\mbf}[1]{\mathbf{#1}}

\numberwithin{equation}{section}

\newcommand{\gM}{\mathcal{M}}
\newcommand{\LL}{\mathcal{L}}	

\newcommand{\gL}{\LL}

\newcommand{\ON}[1]{\mathrm{O}( #1 )}

\newcommand{\SL}[1]{\mathrm{SL}( #1 )}

\usepackage{tikz}
\usetikzlibrary{matrix,arrows,positioning,shapes,snakes,fadings,decorations.pathmorphing,
decorations.pathreplacing,automata,shadows,decorations.markings,patterns}
\usepackage[thicklines]{cancel}
\usetikzlibrary{calc,shapes.callouts,shapes.arrows}

            \usepackage{enumitem}

\begin{document}
\begin{titlepage}

\vfill

\begin{center}
	\baselineskip=16pt  
	
	{\Large \bf  \it Poisson-Lie U-duality in Exceptional Field Theory}
	\vskip 1cm
	{\large \bf Emanuel Malek$^a$\footnote{\tt emanuel.malek@aei.mpg.de}, Daniel C. Thompson$^{b,c}$\footnote{\tt d.c.thompson@swansea.ac.uk}}
	\vskip .6cm
	{\it $^a$ Max-Planck-Institut f\"{u}r Gravitationsphysik (Albert-Einstein-Institut), \\
		Am M\"{u}hlenberg 1, 14476 Potsdam, Germany \\ \ \\
			$^b$ Theoretische Natuurkunde, Vrije Universiteit Brussel, and the International Solvay Institutes, \\ Pleinlaan 2, B-1050 Brussels, Belgium \\ \ \\
			$^c$ Department of Physics, Swansea University, \\ Swansea SA2 8PP, United Kingdom \\ \ \\}
	\vskip 2cm
\end{center}

\begin{abstract}
Poisson-Lie duality provides an algebraic extension of conventional Abelian and non-Abelian target space dualities of string theory and has seen recent applications in constructing quantum group deformations of holography. Here we demonstrate a natural upgrading of Poisson-Lie to the context of M-theory using the tools of exceptional field theory. In particular, we propose how the underlying idea of a Drinfeld double can be generalised to an algebra we call an exceptional Drinfeld algebra. These admit a notion of ``maximally isotropic subalgebras'' and we show how to define a generalised Scherk-Schwarz truncation on the associated group manifold to such a subalgebra. This allows us to define a notion of Poisson-Lie U-duality. Moreover, the closure conditions of the exceptional Drinfeld algebra define natural analogues of the cocycle and co-Jacobi conditions arising in Drinfeld double. We show that upon making a further coboundary restriction to the cocycle that an M-theoretic extension of Yang-Baxter deformations arise. We remark on the application of this construction as a solution-generating technique within supergravity.

\end{abstract}

\vfill

\setcounter{footnote}{0}
\end{titlepage}
 \newpage

\section{Introduction} 

Abelian T-duality asserts in its simplest form that closed strings do not distinguish between circular target space geometries of radius $R$ and $\frac{1}{R}$ measured in units of the string length scale.  This result is a crucial thread of a wider tapestry of U-dualities that are expected to be displayed M-theory whose target spaces contain toroidal directions.  This article will consider some generalised notions of Abelian T-duality and show how they are upgraded to analogous concepts within U-duality.

Non-Abelian T-duality \cite{delaOssa:1992vci} is a proposed dualisation of closed-string non-linear sigma-models (NLSM) whose target space admits the action of non-Abelian isometry group. Whilst the status of Non-Abelian T-duality in terms of the string genus expansion remains unclear, recent evidence \cite{Hoare:2019mcc} at two-loops provides confidence that the duality could remain robust to quantum ($\alpha^\prime$) corrections on the worldsheet. What is absolutely clear is that in the context of holography at large $N$, where both string genus and $\alpha^\prime$ corrections are suppressed, Non-Abelian T-duality can be a powerful solution generating technique as advocated first in \cite{Sfetsos:2010uq,Lozano:2011kb,Lozano:2012au,Itsios:2013wd} (see \cite{Thompson:2019ipl}  for a review and further references) . 

More radically Poisson-Lie (PL) T-duality \cite{Klimcik:1995ux,Klimcik:1995dy} dispenses with the requirement of isometry of a target space but does assume some underlying algebraic structure given by a Drinfeld double.   Here again, though the quantum corrections are far from understood beyond one-loop in $\alpha^\prime$, there have been a number of significant developments.  By exploiting the close connection between PL T-duality and the classical Yang-Baxter equation, this has given rise to wide classes of new integrable NLSMs called $\eta$- or Yang-Baxter sigma-models \cite{Klimcik:2002zj,Klimcik:2008eq}.   Most notably when applied to the $AdS_5\times S^5$ superstring this has led to a marginal deformation that is expected to encode a quantum-group deformation, with $q \in \mathbb{R}$, of ${\cal N}=4$ Super-Yang-Mills gauge theory \cite{Delduc:2013qra,Delduc:2014kha}.  Parallel to this has been the realisation with the $\lambda$-model \cite{Sfetsos:2013wia,Hollowood:2014qma} of an analogous quantum-group deformation  with $q$ a root-of-unity; the $\eta$- and $\lambda$-models are related (at least in simplest bosonic setting where it has been spelt out explicitly) by a PL duality transformation combined with some analytic continuation \cite{Hoare:2015gda,Sfetsos:2015nya,Klimcik:2015gba}.  Alongside $\eta$ and $\lambda$- integrable deformations, non-commutative  $\beta$-deformations realised by TsT transformations can also be thought of as a  Yang-Baxter sigma-model \cite{Matsumoto:2014nra,vanTongeren:2015uha,Osten:2016dvf}. 

Critical to us will be that the most natural understanding of Poisson-Lie T-duality and its associated target spacetimes is provided by the tools and techniques of Double Field Theory (DFT) \cite{Hull:2009mi} and generalised geometry \cite{Gualtieri:2003dx,Hitchin:2004ut,Coimbra:2011nw}. In essence, Poisson-Lie models, i.e. models where Poisson-Lie duality can act, arise as  generalised Scherk-Schwarz reductions \cite{Aldazabal:2011nj,Geissbuhler:2011mx} in which all non-trivial coordinate dependance is encoded in a twist matrix, or generalised frame field, $E_{A}{}^I$ as was first shown in \cite{Hassler:2017yza} and developed in \cite{Demulder:2018lmj}.  Through the Courant bracket, the generalised frame fields of  \cite{Hassler:2017yza} realise the algebra of a Drinfeld double and depend crucially on the properties of the Poisson-Lie bi-vector upon which Poisson-Lie duality relies. When inequivalent twist matrices give rise to the same structure constants of a Drinfeld double (equivalently when there are different splittings of the Drinfeld double into maximally isotropic subalgebras), Poisson-Lie T-duality is realised as an $\ON{d,d}$ duality acting on the Scherk-Schwarz reduced model. Previous attempts to understand Poisson-Lie T-duality and non-Abelian T-duality in DFT and related works can be found in \cite{Catal-Ozer:2019hxw,Sakatani:2019jgu,Bugden:2019vlj}.

An obvious question is then if these generalised Poisson-Lie type T-dualities can be extended to generalised notions of U-duality. Though some partial descriptions have been recently suggested in the literature \cite{Bakhmatov:2019dow}, thus far an algebraically robust description has been lacking.  It is this that we address in the current article in the context of exceptional field theory (ExFT)/exceptional generalised geometry \cite{Pacheco:2008ps,Berman:2010is,Berman:2011cg,Coimbra:2011ky,Berman:2012vc,Coimbra:2012af,Hohm:2013pua}, the M-theoretic analogues to Double Field Theory/generalised geometry. Given the importance of generalised Scherk-Schwarz reductions in the DFT realisation of Poisson-Lie T-duality, it is natural to use the analogous ExFT structure \cite{Berman:2012uy,Lee:2014mla,Hohm:2014qga} to build a notion of Poisson-Lie U-duality. This is our goal in this paper.

In ExFT, a splitting is made of $11-d$ ``external'' directions and $d$ ``internal'' directions such that all field content and gauge symmetries are repackaged into representations of the exceptional Lie groups $E_{d(d)}$. Our focus will be case of $d=4$ such that the duality group is $SL(5)$ and we will restrict our attention to the ``internal'' directions in which the Poisson-Lie U-duality will act.

We begin with a review of Poisson-Lie T-duality and its realisation in Double Field Theory in \ref{s:PLReview}, before  reviewing the relevant $SL(5)$ Exceptional Field Theory in section \ref{s:SL5Review} and  introducing the natural analogue of a Drinfeld Double, which for want of a better name we call an Exceptional Drinfeld Algebra in section \ref{s:Drinfeld}. In section \ref{s:Frame} we will show how a generalised frame can be introduced that reproduces the algebra of the Exceptional Drinfeld Algebra, allowing us to define a U-duality notion of Poisson-Lie duality. Finally, in sections \ref{s:QC} and \ref{s:Coboundary} we develop methods for constructing an Exceptional Drinfeld Algebra starting from a Lie Group $G$, and thereby also suggest a U-dual version of Yang-Baxter deformations. Finally, we present some examples in section \ref{s:Examples} and conclude in section \ref{s:Conclusions} with a brief outlook.

\paragraph{Note added:} While finalising this manuscript, the paper \cite{Sakatani:2019zrs} appeared which proposes a U-duality extension of Drinfeld Doubles and has some overlap with our sections \ref{s:Drinfeld} and \ref{s:Frame} in the case where our $I_a = \tau_{a5}$.

\section{Review of Poisson-Lie T-duality} \label{s:PLReview}

In this section we provide a brief recap of Poisson-Lie T-duality. We will flip the conventional exposition by starting with algebraic considerations to eventually arrive at an associated NLSM describing the NS sector of a closed string; this  will serve as a road map for what follows.  

Central to the construction is a (classical) Drinfeld double: an even-dimensional real algebra $\frak{d}$ with generators $T_A$ obeying $[T_A, T_B]= i F_{AB}{}^CT_C$ equipped with a symmetric split-signature ad-invariant pairing $\eta(\cdot , \cdot)$ such that $\frak{d}$ admits at least one decomposition $\frak{d} = \frak{g} \oplus \tilde{\frak{g}}$ with   $\frak{g}$ and $\tilde{\frak{g}}$   sub-algebras that are maximally isotropic with respect to $\eta$.    Letting $t_a$ ($\tilde{t}^a$) be generators for $\frak{g}$  ($\tilde{\frak{g}}$)    we have
 \begin{equation}
 \begin{aligned}\label{eq:doublealgebra}
 & \eta(t_a ,t_b ) = 0 \ , \quad   \eta(t_a ,\tilde{t}^b ) =  \delta_a^b \ ,  \quad \eta(\tilde{t}^a , \tilde{t}^b) = 0  \ ,  \\
  &  [t_a , t_b]= i f_{ab}{}^c t_c\ , \quad  [t_a , \tilde{t}^b]= i \tilde{f}^{b c }{}_a t_c -  i f_{ac}{}^b \tilde{t}_c  \ , \quad [\tilde{t}^a , \tilde{t}^b]= i \tilde{f}^{ab}{}_c\tilde{t}^c \, .
 \end{aligned}
 \end{equation}
The Jacobi identity for $\frak{d}$ imposes a compatibility constraint, 
\be\label{eq:cocy}
0 = 2 \tilde{f}^{ed}{}_{[a} f_{b]d}{}^c - 2 \tilde{f}^{cd}{}_{[a} f_{b]d}{}^e -  \tilde{f}^{ec}{}_d f_{ab}{}^d \,,
\ee
which is better understood as demanding that $\delta(t_a) = \tilde{f}^{bc}{}_a t_b \otimes t_c$ defines a one-cocycle on $\frak{g}$ valued in $  \frak{g}\wedge \frak{g}$ which, as a trivial consequence of the Jacobi identity for $\tilde{\frak{g}}$, also obeys a co-Jacobi identity.  Equivalent we may then speak of $(\frak{g}, \delta)$ as defining a Lie bi-algebra. 

The exponential of a Lie bi-algebra is a Poisson-Lie group, that is a Lie-group manifold $G$ equipped with a Poisson bi-vector  compatible with the group composition law and obeying the Schouten identity. Equivalent to this Poisson bi-vector is a one-cocyle on $G$ valued in $\frak{g}\wedge \frak{g}$ denoted by $\Pi = \Pi_g^{ab} t_{a} \otimes t_b$  which is constructed from the adjoint action of  $g \in G$ on $\frak{d}$ as follows: 
\begin{eqnarray}
g \cdot t_a \cdot g^{-1}  =   (a_{g})_a{}^b t_b \ , \quad g \cdot \tilde{t}^a \cdot g^{-1}  =   (b_g)^{ab}   t_b  + ( a_{g^{\tiny{-1}}} )_b{}^a \tilde{t}^b \, , \quad
\Pi_{g}^{ab} =  (b_g)^{ac}( a_{g^{\tiny{-1}}} )_c{}^b   \, .
\end{eqnarray}
As a consequence of this definition,   $\Pi$ enjoys some useful properties including:
\be\label{eq:Piproperties} 
\begin{aligned}
\Pi_{g}^{ab} =  (b_g)^{ac}( a_{g^{\tiny{-1}}} )_c{}^b = - \Pi_{g}^{ba}\, ,  \\
\Pi_{hg} = \Pi_{g} + (a_{g^{\tiny{-1}}} \otimes a_{g^{\tiny{-1}}} ) \Pi_h \, , \quad \Pi_e = 0 \, , \\
d\Pi_{g}^{ab} =  - l^c \tilde{f}^{ab}{}_c - 2 l^c f_{cd}{}^{[a}\Pi^{b] d} \, , 
\end{aligned}
\ee
in which we have introduced the left-invariant one-forms   $l = i l^a t_a  =  g^{-1} d g$. The dual vector fields to $l^a$ will be denoted $v_a$. It will be useful in what follows to build from $\Pi^{ab}$ a second set of vector fields $\pi^a \equiv \Pi_g^{ab} v_b $.  A modest calculation, appealing to the properties of eq.~\eqref{eq:Piproperties},  shows that these objects realise the algebra of $\frak{d}$ given in eq.~\eqref{eq:doublealgebra} under the Lie derivative
 \be
 \begin{aligned}
 L_{v_a} v_b = - f_{ab}{}^c v_c \, ,  \quad  L_{v_a}\pi^b = f_{ac}{}^b \pi^c - \tilde{f}^{bc}{}_a v_c \, , \quad L_{\pi^a} \pi^b = - \tilde{f}^{ab}{}_c \pi^c  \,   . 
 \end{aligned}
 \ee 
 
 We now upgrade this discussion to generalised geometry which concerns the generalised tangent bundle $E$ given locally as $TG\oplus T^\star G$.  On $E$ there is a generalised Lie (Dorfman) derivative which acts on two sections $U = u^i \partial_i + \mu_i dx^i$ and $V = v^i \partial_i + \nu_i dx^i$   by 
 \be\label{eq:Dorfman}
 {\cal L}_U V = L_u v + \left( L_u \nu - \iota_v d \mu \right) \, . 
 \ee
 In particular we can introduce two sets of generalised vectors, 
 \be\label{eq:PLgenframe}
 E_a = v_a \ , \quad \tilde{E}^a = \pi^a + l^a \, , 
 \ee
 which we package together  by defining a set of {\em generalised frame fields}  $E_A = (E_a, \tilde{E}^a)$ that under the Dorfman derivative furnish the algebra of the double $\frak{d}$,  
 \be 
 {\cal L}_{E_A} E_B = -  F_{AB}{}^C E_C \ . 
 \ee   
 
We may now perform a generalised Scherk-Schwarz reduction of type II supergravity on $G$. For this, introduce a set of  $d^2$ constants $E_0 = G_0 + B_0$ assembled into constant {\em generalised metric}, ie. a representative   of the coset $\ON{d,d}/\ON{d} \times \ON{d}$ given by
\be
 \mathring{\cal H}_{AB}  = \begin{pmatrix} G_0 - B_0 G_0^{-1} B_0 & - B_0 G_0^{-1}  \\ G^{-1}_0 B_0  & G^{-1}_0 \end{pmatrix} \, , \quad  \mathring{\cal H}_{AB} = \eta_{AC} \mathring{\cal H}^{CD}\eta_{DB} \ , \quad \eta_{AB} = \eta(T_A, T_B) \, .  
\ee
 We can use the generalised frame field to translate this constant   {\em generalised metric} to  one defined on $E\approx TG\oplus T^\star G$ according to 
 \be
 {\cal H} =  \mathring{\cal H}^{AB} E_A \otimes E_B  \, . 
 \ee
 
From this ``curved" generalised metric, we can extract the metric tensor and B-field $E_{ij}= G_{ij}+ B_{ij}$ on $G$ which match the target space geometry of the Poisson-Lie $\sigma$-model with action \cite{Klimcik:1995dy}
\be\label{eq:actS}
  S = \int d^2 \sigma \,~ l_+^a \left[ \left(E_0^{-1} + \Pi_g   \right)^{-1} \right]_{ab} l_-^b \ , 
\ee
 in which the left invariant one-forms on $G$ have been pulled back to the worldsheet and $\pm$ signify light-cone coordinates.  The would-be-Noether currents,  ${\cal J}_{\pm a} = v_a^i (G_{ij} \pm B_{ij}) \partial_\pm x^i$,  associated to the $G$ action generated by the vector fields $v_a$ are not conserved but enjoy a remarkable non-commutative conservation law, 
\be
  d \star {\cal J}_a = \tilde{f}^{bc}{}_a \star {\cal J}_b \wedge \star {\cal J}_c \, , 
\ee
that is often called a Poisson-Lie symmetry. 
   
We could of course swap the role of $\frak{g}$ and $\tilde{\frak{g}}$ in the entire discussion above constructing $\tilde{\Pi}_{ab}, \tilde{\pi}_a, \tilde{v}^a, \tilde{l}_a$ as well as generalised frame fields $\tilde{E}_A$.  This results in a dual $\sigma$-model, $\tilde{S}$, defined on $\tilde{G}= \exp\tilde{\frak{g}}$ that is canonically equivalent to the first \cite{Klimcik:1995dy,Sfetsos:1996xj,Sfetsos:1997pi}. A set of Buscher rules for such a dualisation is easily formulated in the generalised geometry by  starting with a curved space generalised metric $\cal{ H}$, undressing the frame fields $E_A$ to return to the flat space generalised metric $\mathring{\cal H}$, performing an $O(d,d)$ action that implements the swapping $(T_a, \tilde{T}^a) \leftrightarrow  ( \tilde{T}^a, T_a)$ and then re-dressing with the generalised frames  $\tilde{E}_A$.

The action in eq.\eqref{eq:actS} is one-loop renomalisable \cite{Valent:2009nv} and the RG equations governing the two sigma models $S$ and $\tilde{S}$ are  are equivalent at one-loop  \cite{Sfetsos:2009dj}  and can be formulated in terms of a renormalisation to $\mathring{\cal H}$ \cite{Avramis:2009xi,Sfetsos:2009vt}.  Should the target space of the original theory define a solution of the appropriate (super)gravity (or part thereof) under normal circumstances so too will the dual, and hence this procedure defines a solution generating technique.  This is called Poisson-Lie T-duality.

\section{Brief Review of $\SL{5}$ Exceptional Field Theory} \label{s:SL5Review}
In the ExFT approach to eleven-dimensional supergravity a split is made into a $d$ dimensional ``internal'' space $M$  and $11-d$ ``external'' directions but importantly no restriction is made on the coordinate dependence of any fields and no truncation is assumed from the outset. This splitting enables a rewriting of the variables of supergravity in a way that makes manifest the $E_{d(d)}$ U-duality symmetry. In order to get efficiently to the core issues, our focus in this paper will be exclusively on the internal directions and we shall ignore both dependance of internal fields on external coordinates and external fields entirely -- these extra modes will   be spectators as far as the Poisson-Lie U-duality is concerned.  Our approach is essentially a specialisation of the general construction of gauged supergravities via ExFT \cite{Berman:2012uy,Lee:2014mla,Hohm:2014qga,Inverso:2017lrz} and the inclusions of these spectator modes is well discussed in the literature.

On the internal space the bosonic field content, namely the components of the metric tensor and three-form and dual six-form potentials, are packaged into a generalised metric ${\cal M}_{AB}$ where the index $A$ runs over a particular representation $R$ of the duality group \cite{Hull:2007zu,Pacheco:2008ps}. $R$ is sometimes called the coordinate representations\footnote{In ExFT it is sometimes asserted that space is augmented by extra coordinates to form a multiplet $X^A$ in the representation $R$ -- however, the price paid is that additional constraints are required to reduce the dynamics to depend only on the conventional $d$ coordinates (or fewer).  Here we will only consider objects that depend on $d$ coordinates and need not invoke this. } and the sections of its associated fibre bundle, the generalised tangent bundle $E$, are called generalised vector fields and generate diffeomorphisms, two-form and 5-form gauge transformations on the internal space $M$ via the generalised Lie derivative.

As it is most amenable for direct calculation, we will henceforth consider the case of $d=4$. In this case, $E_{4(4)} \cong SL(5)$ and the representation $R =\bf{10}$ and we will use the composite index notation $A=[ {\cal A} \cal{B}]$ with ${\cal A}= 1\dots 5$. For $d=4$, the generalised tangent bundle $E\approx T M \oplus \wedge^2 T^\star M$ consists of vector fields and two-forms, generating diffeomorphisms and gauge transformations on $M$. The generalised Lie derivative acts on generalised vector fields as introduced in eq.~\eqref{eq:Dorfman} with the replacement of one-forms to two-forms. For practical reasons, we will often prefer to work with the fundamental $\bf{5}$ representation of $SL(5)$, with associated bundle $ \Lambda^0 T^\star M \oplus \wedge^3 T^\star M$, the sections of which undergo diffeomorphisms and gauge transformations mediated by the generalised Lie derivative as follows
\begin{equation}\label{eq:Lie5}
{\cal L}_{U} W=  L_u W^{(0)} + \left(L_u W^{(3)} +  W^{(0)} d\mu \right) \,,
\end{equation} 
where the generalised vector field $U = u + \mu$, $W = W^{(0)} + W^{(3)}$ and $u$ is a vector field, $\mu$ is a 2-form, $W^{(0)}$ is a scalar field and $W^{(3)}$ is a 3-form.

\section{Exceptional Drinfeld Algebra} \label{s:Drinfeld}
Recall that a Drinfeld double, $\mathfrak{d}$, is an algebra that (i) is equipped with a symmetric split-signature ad-invariant pairing $\eta(\cdot,\cdot)$ and (ii) admits at least one decomposition into maximally isotropic subalgebras $\mathfrak{g}$, $\tilde{\mathfrak{g}}$ with respect to $\eta$ such that $\mathfrak{d} = \mathfrak{g} \oplus \tilde{\mathfrak{g}}$. Let us now construct the analogue of a Drinfeld double which, for want of inspiration, we call  an ``Exceptional Drinfeld Algebra'' and denote by $\eda$. We will do this by generalising the above notions from $O(d,d)$ to $SL(5)$, mirroring the generalisation of generalised geometry or gauged supergravities from $O(d,d)$ to $SL(5)$. Thus, we define (in the specific case of $SL(5)$)  an Exceptional Drinfeld Algebra, $\eda$, as a (i) $\left(n \leq 10\right)$-dimensional subalgebra of $\mathfrak{sl}(5) \oplus \mathbb{R}^+$, which contains at least one (ii) 4-dimensional subalgebra $\mathfrak{g}$, that satisfies the ``maximal isotropy'' condition for $SL(5)$ which we will now define.

Let us write the generators of $\eda$ in a 10-dimensional $\SL{5}$ covariant manner, such that they are represented by
\begin{equation}
 \eda = \mathrm{span}\left( T_{\cA\cB} \right) \,,
\end{equation}
with $T_{\cA\cB} = T_{[\cA\cB]}$ and $\cA, \cB = 1, \ldots, 5$. However, as we will make explicit later, typically $n < 10$  so the generators $T_{\cA\cB}$ are not all linearly independent. Since $\eda$ is a subalgebra of $\mathfrak{sl}(5) \oplus \mathbb{R}^+$, we can introduce the $\mathfrak{sl}(5)$-invariant $\epsilon_{\cA\cB\cC\CD\cE}$ which is left invariant up to scalings.

The ``maximal isotropy'' condition for the subalgebra $\mathfrak{g}$ is that its generators satisfy
\begin{equation} \label{eq:MaxIso}
\epsilon^{\cA\cB\cC\CD\cE} T_{\cA\cB} T_{\cC\CD} = 0 \,,
\end{equation}
where this should be understood as a relation on the universal enveloping algebra. We can always label these generators as $t_a \equiv T_{a5}$ with  $a=1\dots 4$ and the remaining 6 generators by $\tilde{t}^{ab} \equiv \frac{1}{2} \epsilon^{ab cd} T_{cd} $.

Finally, let us discuss the structure constants of $\eda$. In analogy to 7-d gauged maximal SUGRA \cite{Samtleben:2005bp}, $\eda$ is a Leibniz algebra with structure constants given by
\begin{equation}\label{eq:drinfeldext}
 \left[ T_{\cA\cB},\, T_{\cC\CD} \right] = \frac{i}{2}\, F_{\cA\cB,\cC\CD}{}^{\cE\cF} T_{\cE\cF} \,,
\end{equation}
where the structure constants, also called the embedding tensor, are given by
\begin{equation}
 \begin{split} \label{eq:StrConst10}
  F_{\cA\cB,\cC\CD}{}^{\cE\cF} &= 4 F_{\cA\cB,[\cC}{}^{[\cE} \delta^{\cF]}_{\CD]} \,, \\
  F_{\cA\cB,\cC}{}^{\CD} &= \frac12 \epsilon_{\cA\cB\cC\cE\cF} Z^{\cE\cF,\CD} + \frac12 \delta_{[\cA}^{\CD} S_{\cB]\cC} + \frac13 \delta^{\CD}_{[\cA} \tau_{\cB]\cC} + \frac16 \delta^{\CD}_{\cC} \tau_{\cA\cB} \,.
 \end{split}
 \end{equation}
 Here $S_{\cA\cB}$ is symmetric, $\tau_{\cA\cB}$ is antisymmetric and $T_{\cA\cB\cC}{}^{\cal D} = \frac1{2} \epsilon_{\cA\cB\cC\cal E\cal F} Z^{\cal E \cal F,\cal D}$ is traceless, i.e. $ 
 Z^{[\cA\cB,\cC]} = 0$. 
This is, in general, not a Lie algebra because the structure constants need not be anti-symmetric under the interchange of the two sets of lower indices. Moreover, instead of the Jacobi identity, closure of $\eda$ requires the quadratic constraint (also known as the Leibniz identity)
\begin{equation} \label{eq:QC}
2 F_{\cA\cB,[\cC}{}^{\cal G} F_{|{\cal G}|\CD],\cE}{}^{\cF} - F_{\cA\cB,\cal G}{}^{\cF} F_{\cC\CD,\cE}{}^{\cal G} + F_{\cA\cB,\cE}{}^{\cal G} F_{\cC\CD,\cal G}{}^{\cF} = 0 \,. 
\end{equation}
 
Requiring closure under the adjoint action of $g\in G$ on $\eda$ implies that
\begin{equation} \label{eq:AdjGd}
 g \cdot T_{\cA\cB} \cdot  g^{-1}  = \frac{1}{2} (A_g)_{ \cA\cB}{}^{\cC\CD} T_{\cC\CD} \,.
\end{equation}
Moreover, since $\eda \subset \mathfrak{sl}(5) \oplus \mathbb{R}^+$, this adjoint action must lie inside $\SL{5} \times \mathbb{R}^+$, i.e.
\begin{equation}
 (A_g)_{ \cA\cB}{}^{\cC\CD} =2 (A_g)_{ [\cA}{}^{[\cC}  (A_g)_{  \cB]}{}^{ \CD]} \,.
\end{equation}
Finally, since $\mathfrak{g} \subset \eda$ is a subalgebra, we must have
\begin{equation}
 g \cdot t_a \cdot g^{-1} = (ad_{g})_a{}^b\, t_b \,,
\end{equation}
where $ad_g$ denotes the adjoint action of $g \in G$ on $\mathfrak{g}$. As a result, the $\SL{5} \times \mathbb{R}^+$ matrix $\left(A_g\right)_{\cA}{}^{\cB}$ is given by
\begin{equation} \label{eq:AdjGdlambda}
\left(A_g\right)_{\cA}{}^{\cB} = \begin{pmatrix}
\Xi_g^{1/3} |ad_g|^{-1/3} (ad_g)_a{}^b & \Xi_g^{-1/3} |ad_g|^{1/3} (\lambda_g)_a \\
0 & \Xi_g^{-1/3} |ad_g|^{1/3}
\end{pmatrix} \,,
\end{equation}
where $|ad_g| = \det\left(ad_g\right)$ is the determinant of the adjoint action of $g \in G$, $\Xi_g = \det\left(A_g\right)$ is the overall $\mathbb{R}^+$ action and $(\lambda_g)_a$ defines the action of $g \in G$ on $\tilde{t}^{ab}$ as
\begin{equation}
 g \cdot \tilde{t}^{ab} \cdot g^{-1} = \Xi_g^{2/3} |ad_g|^{1/3} (ad_{g^{-1}})_c{}^a (ad_{g^{-1}})_d{}^b\, \tilde{t}^{cd} + \lambda^{abc}\, (ad_g)_c{}^e\, t_e \,,
\end{equation}
where $(\lambda_g)^{abc} = \epsilon^{abcd} (\lambda_g)_d$.

Equations \eqref{eq:AdjGd} and \eqref{eq:AdjGdlambda} imply the following properties for $(\lambda_g)_a$:
\begin{itemize}
	\item $(\lambda_g)_a$ vanishes at the identity of $G$
	\begin{equation}
	 (\lambda_{e})_a = 0 \,. \label{eq:lambda0}
	\end{equation}
	\item $(\lambda_g)_a$ inherits a group composition rule from that of the adjoint action 
	\begin{equation}\label{eq:groupcomp}
	(\lambda_{hg})_a = (\Xi_g)^\frac{2}{3} |ad_g|^{-\frac{2}{3}} (ad_g)_a{}^b (\lambda_h)_b + (\lambda_g)_a \,,
	\end{equation}
	which is reminiscent of the composition law obeyed by the Poisson-Lie bi-vector.  
\end{itemize}
In the following, we will often drop the subscript $g$ on $\Xi_g$ and $(\lambda_g)_a$ for simplicity. 
 
The structure constants of the algebra $\eda$ can now be related to $\lambda_a$ using
\begin{equation} \label{eq:dTrick}
 d \left( g\, T_{\cA\cB}\, g^{-1} \right) = i\, l^c g \left[ t_c,\, T_{\cA\cB} \right] g^{-1} \,,
\end{equation}
where $l = i\,l^a\,t_a = g^{-1}\,dg$ denotes the Maurer-Cartan one-form as in section \ref{s:PLReview}. We first use that $\mathfrak{g}$ is a subalgebra of $\eda$ with structure constants $f_{ab}{}^c$, which we decompose into a traceless part $\hat{f}_{ab}{}^c$ and trace $I_a$, given by
\begin{equation}
f_{ab}{}^c = \hat{f}_{ab}{}^c + \frac23 \delta^c_{[b} I_{a]} \,, \quad I_a = f_{ab}{}^b \,.
\end{equation}
This allows us to identify
\begin{equation} \label{eq:EmbgAlg}
 S_{a5} = \frac43 I_a + \frac23 \tau_{a5} \,, \qquad T_{ab5}{}^c = \hat{f}_{ab}{}^c \,, \qquad S_{55} = 0 \,.
\end{equation}
We then find from \eqref{eq:dTrick}
\begin{equation}
 \begin{split} \label{eq:dlambda}
  d\Xi &= \frac12 l^a\, \tau_{a5}\, \Xi \,, \\
  \lambda_{(a,b)} &= - \frac14 S_{ab} + \frac23 \lambda_{(a} I_{b)} + \frac13 \lambda_{(a} \tau_{b)5} \,, \\
  \lambda_{[a,b]} &= \frac16 \tau_{ab} + T_{5ab}{}^5 + \hat{f}_{ab}{}^c \lambda_c + \frac13 \lambda_{[a} \tau_{b]5} \,,
 \end{split}
\end{equation}
where we defined the derivative of $\lambda_{a}$ as 
\begin{equation}
 d\lambda_{a} = \lambda_{a,b}\, l^b \,.
\end{equation}

Moreover, we can use invariance of the structure constants of $\eda$ under the adjoint action of $G$. This implies
\begin{equation} \label{eq:AdInv}
 F_{\cA\cB,\,\cC\CD}{}^{\cal GH} A_{\cal G}{}^{\cE} A_{\cal H}{}^{\cF} = A_{\cA}{}^{\cal G} A_{\cB}{}^{\cal H} A_{\cC}{}^{\cal I} A_{\CD}{}^{\cal J} F_{\cal G H, I J}{}^{\cE\cF} \,,
\end{equation}
and thus various relations between the structure constants \eqref{eq:StrConst10}, $\Xi_g$, $ad_g$ and $(\lambda_g)_a$ which will be important for us in the following.
 
\section{The Frame Fields and Embedding Tensor} \label{s:Frame}
The next step is to furnish the algebra eq. \eqref{eq:drinfeldext} via the generalised Lie derivative acting on a set of generalised frame fields.
By analogy with the Poisson-Lie case of eq. \eqref{eq:PLgenframe} we will construct the generalised frame fields out of objects on $\eda$, in particular $\lambda^{abc}$, as follows
\begin{equation} \label{eq:Frame}
 E_a = v_a \ , \quad \tilde{E}^{ab} = \lambda^{abc}\, v_c  +  \alpha\, l^a \wedge l^b \,,
\end{equation}
where  
$\lambda^{abc} = \epsilon^{abcd} \lambda_d$, $\alpha$ is a function on $G$ and $v_a$ and $l^a$ are the left-invariant vector fields and one-forms on $G$.

Taken together these define generalised frame fields in the ${\bf 10}$, ie. $E_{\cA  \cB} = ( E_{a}, \frac{1}{2} \epsilon_{ab cd} \tilde{E}^{cd})$, that should obey
\be  \label{eq:FrameDif}
 {\cal L}_{E_{\cA\cB}} E_{\cC\CD} = -   F_{\cA\cB,\cC\CD}{}^{\cE\cF} E_{\cE\cF} \,,
\ee
with $F_{\cA\cB,\cC\CD}{}^{\cE\cF}$ given in \eqref{eq:StrConst10}.

In addition the frame fields must define $\SL{5}\times \mathbb{R}^+$ group elements and so can be decomposed in terms of objects in the $\mathbf{\bar{5}}$ representation:   
\begin{equation}
 E_{\cA \cB}{}  = 2 \Delta^{-1} {\cal E}_{[\cA}{}  {\cal E}_{\cB]}{}  \,,
\end{equation}
where under the generalised Lie-derivative $\Delta$ has weight 1 \footnote{Here $\cE_{\cA}{}^I$ has the natural weight $\frac35$.}. It is often easier to work with the ${\cal E}_{a}$, which are given by
\begin{equation}
 {\cal E}_{a}  = \frac1{3!} \epsilon_{abcd} \left( \alpha^2\, l^b \wedge l^c \wedge l^d - \alpha\, \lambda^{bcd} \right) \,, \quad 
 {\cal E}_5  = \alpha \,,  \quad  \Delta = \alpha^3\, |l|  \,,
\end{equation}
and must obey
\begin{equation}\label{eq:genLieon5}
 \gL_{E_{\cA\cB}} {\cal E}_{\cC} = F_{\cA \cB, \cC}{}^{\cal D}  {\cal E}_{\cal D} + \frac12 \tau_{\cA\cB} {\cal E}_{\cC} \,,
\end{equation}
in which we recall the structure constants (embedding tensor)  in the fundamental defined in eq.~\eqref{eq:StrConst10}.

We will now show that, with the $\lambda_a$ defined in section \ref{s:Drinfeld}, via equation \eqref{eq:AdjGdlambda}, and an appropriate function $\alpha$ that we will specify shortly, the $E_{\cA\cB}$ of \eqref{eq:Frame} satisfy the differential conditions \eqref{eq:FrameDif}, once certain further restrictions are placed on the structure constants \eqref{eq:StrConst10}.  Firstly, we evaluate \eqref{eq:genLieon5} and find: 
\begin{equation} \label{eq:LinCon}
 \begin{aligned}    S_{a5} = 2 L_{v_a} \ln \alpha + 2 I_a \, , \quad  S_{55} = 0  \, ,  \quad S_{ab} = 4 I_{(a} \lambda_{b)} - 4 \lambda_{(a,b)} + 4 \lambda_{(a} L_{v_{b)}} \ln \alpha \, , \\   
 \tau_{a 5} = 3 L_{v_a} \ln \alpha + I_a \, , \quad  \tau_{ab} = - 2 \left( 3 \lambda_{[a} L_{v_{b]}} \ln \alpha + \lambda_{[a} I_{b]} + \lambda_{[a,b]} \right) \, ,\\
 T_{ab5}{}^c= \hat{f}_{ab}{}^c \, , \quad T_{ab5}{}^5 = - \hat{f}_{ab}{}^c\lambda_c + \frac43 \lambda_{[a,b]}  \, , \\
 T_{abc}{}^d =   3   \hat{f}_{[ab}{}^d\lambda_{c]} - 2 \lambda_{[a,b} \delta_{c]}^d  \, , \quad  T_{abc}{}^5 =     -3    \hat{f}_{[ab}{}^d\lambda_{c]} \lambda_d  + 6 \lambda_{[a,b} \lambda_{c]} \, .
 \end{aligned}  
\end{equation}
Recall  from \eqref{eq:lambda0} that $\lambda_a$ vanishes at the identity of $G$. Since the expressions \eqref{eq:LinCon} define structure constants, we can evaluate them at the identity where $\lambda_a = 0$. This implies the following relation between the structure constants.
\begin{equation}\label{eq:Ttau}
\begin{split}
 T_{abc}{}^5 = 0 \,, \qquad T_{ab5}{}^{5}= -\frac{2}{3} \tau_{ab} \,, \qquad T_{abc}{}^d = \tau_{[ab}\delta_{c]}^d\,.
\end{split}
\end{equation}
Therefore, to perform a Poisson-Lie U-duality, we must impose the further restriction on the algebra $\eda$ that its structure constants \eqref{eq:StrConst10} must satisfy \eqref{eq:Ttau}.

With the structure constants related by \eqref{eq:Ttau}, we can derive the following three equations from the adjoint-invariance condition of the structure constants \eqref{eq:AdInv}.
\begin{equation}
\begin{split} \label{eq:constancyconditions}
0 &= f_{ab}{}^c \lambda_c +  \lambda_{[a} \tau_{b]5} +\frac{2}{3} \lambda_{[a} K_{b]} \,, \\
0 &= f_{[ab}{}^d \lambda_{c]} -  f_{[ab}{}^e \delta_{c]}^d \lambda_e  - \frac{4}{3} \lambda_{[a} K_{b} \delta_{c]}^d \,, \\
0 &= \tau_{[ab}\lambda_{c]} \,,
\end{split}
\end{equation}
where $K_a = I_a + \frac12 \tau_{a5}$. These relations together with \eqref{eq:dlambda} and for the structure constants related by \eqref{eq:Ttau} imply that the $\lambda_a$ introduced in \eqref{eq:AdjGdlambda} and
\begin{equation} \label{eq:alpha}
 \alpha_g = \alpha_0 (\Xi_g)^{2/3} |ad_g|^{1/3} \,,
\end{equation}
for some constant $\alpha_0$, satisfy precisely the right differential conditions, i.e.
\begin{equation} \label{eq:allfluxes}
\begin{aligned}
S_{a5} = 2 L_{v_a} \ln \alpha + 2 I_a \, , \quad  S_{55} = 0  \, ,  \quad S_{ab} = 4 I_{(a} \lambda_{b)} - 4 \lambda_{(a,b)} + 4 \lambda_{(a} L_{v_{b)}} \ln \alpha \, , \\   
\tau_{a 5} = 3 L_{v_a} \ln \alpha + I_a \, , \quad  \tau_{ab} = - 2 \left( 3 \lambda_{[a} L_{v_{b]}} \ln \alpha + \lambda_{[a} I_{b]} + \lambda_{[a,b]} \right)\, ,  \\
T_{ab5}{}^c= \hat{f}_{ab}{}^c \, , \quad -\frac23 \tau_{ab} = - \hat{f}_{ab}{}^c\lambda_c + \frac43 \lambda_{[a,b]}  \, , \\
\tau_{[ab} \delta_{c]}^d =   3   \hat{f}_{[ab}{}^d\lambda_{c]} - 2 \lambda_{[a,b} \delta_{c]}^d  \, , \quad  0 = -3    \hat{f}_{[ab}{}^d\lambda_{c]} \lambda_d  + 6 \lambda_{[a,b} \lambda_{c]} \,.
\end{aligned}  
\end{equation}
Therefore, the generalised Lie derivative of $E_{\cA\cB}$ given in \eqref{eq:Frame} furnish the algebra of the Drinfeld extension \eqref{eq:StrConst10} when the structure constants are related by \eqref{eq:Ttau}.

With the relations \eqref{eq:Ttau}, the algebra relations of the Exceptional Drinfeld algebra reduce to
\begin{equation} \label{eq:AlgRel}
\begin{split}
\left[ T_{a5},\, T_{b5} \right] &= i\, f_{ab}{}^c\, T_{c5} \,, \\
\left[ T_{a5},\, T_{bc} \right] &= - \left[ T_{bc},\, T_{a5} \right] =i \left( 2\, f_{a[b}{}^d T_{c]d} +   \frac23 K_a\, T_{bc}  + \frac12 \epsilon_{bcde}\, \tilde{f}^{def}{}_{a} T_{f5} \right) \,, \\
\left[ T_{ab},\, T_{cd} \right] &= \frac{i}{3} \left( \tilde{f}^{efg}{}_b\, \epsilon_{efg[c}\, T_{d]a} - \tilde{f}^{efg}{}_a\, \epsilon_{efg[c}\, T_{d]b} \right) + \frac{i}{2} \epsilon_{abef}\, \tilde{f}^{efg}{}_g\, T_{cd}  \,.
\end{split}
\end{equation}
Here we have introduced the ``dual structure constants'' $\tilde{f}^{abc}{}_d$ given by
\begin{equation} \label{eq:DualStrConst0}
 \tilde{f}^{abc}{}_d = -4\, \epsilon^{abce} \left(S_{de} - 2\, \tau_{de} \right) \, ,
\end{equation}
and we also make use of the definitions  
\begin{equation}
K_a =  I_a + \frac{1}{2} \tau_{a5} \, , \quad L_a =  \tau_{a5}- I_a . 
\end{equation}
In terms of the generators $\tilde{t}^{ab}$, the Exceptional Drinfeld algebra relations \eqref{eq:AlgRel} take the form
\begin{equation}
 \begin{split}
  \left[ t_a,\, t_b \right] &= i\, f_{ab}{}^c\, t_c \,, \\
  \left[ t_{a},\, \tilde{t}^{bc} \right] &= - \left[ \tilde{t}^{bc},\, t_{a} \right] =i \left(- 2\, f_{ad}{}^{[b}\, \tilde{t}^{c]d} + \frac13 L_a\, \tilde{t}^{bc} + \tilde{f}^{bcd}{}_a\, t_d \right) \,, \\
  \left[ \tilde{t}^{ab},\, \tilde{t}^{cd} \right] &= i \left(-2\, \tilde{f}^{cd[a}{}_e\, \tilde{t}^{b]e} + 2\, \tilde{f}^{cde}{}_e\, \tilde{t}^{ab} + \tilde{f}^{abe}{}_e\, \tilde{t}^{cd} \right) \,.
 \end{split}
\end{equation}
Note that the Exceptional Drinfeld Algebra is a Leibniz algebra and the $\left[ \tilde{t}^{ab}, \tilde{t}^{cd} \right]$ is not antisymmetric.

With the information above we can also give the derivative of $\lambda^{abc}$ in a form that mirrors that displayed by the PL bivector $\Pi$ \eqref{eq:Piproperties} namely:
\begin{equation}
d \lambda^{abc} =  \tilde{f}^{abc}{}_d\, l^d  + 3 f_{e d}{}^{[a}\, \lambda^{bc]d}\, l^e +\frac{1}{3} \lambda^{abc}\, L_e\, l^e  \,. 
\end{equation}

\subsection{Poisson-Lie U-dualities}
When an exceptional Drinfeld algebra admits two different subalgebras $\mathfrak{g}$ and $\tilde{\mathfrak{g}}$ satisfying the ``maximal isotropy conditions'' \eqref{eq:MaxIso}, we can perform a Poisson-Lie U-duality as follows. Firstly, note that the generators of $\mathfrak{g}$ and $\tilde{\mathfrak{g}}$ are necessarily related by some $\SL{5}$ transformation ${\cal T}_{\cA}{}^{\cB}$. Next, for both $\mathfrak{g}$ and $\tilde{\mathfrak{g}}$, we can introduce $\Xi$, $\lambda_a$ and $\Xi'$, $\lambda'_a$, respectively as shown in \ref{s:Drinfeld}. We then construct the frame fields $\cE_{\cA}$ and $\cE'_{\cA}$ realising the algebra of the exceptional Drinfeld algebra based on $\mathfrak{g}$, $\tilde{\mathfrak{g}}$, respectively, using our results in \ref{s:Frame}. Finally, we can perform a generalised Scherk-Schwarz reduction of 11-dimensional supergravity on $G$ using the Ansatz
\begin{equation}
 \gM_{\cA\cB} = \cE_{\cA}{}^{\bar{\cA}} \cE_{\cB}{}^{\bar{\cB}} \gM_{\bar{\cA}\bar{\cB}} \,,
\end{equation}
where $\gM_{\bar{\cA}\bar{\cB}}$ are the scalar fields of 7-dimensional maximal gauged supergravity and are thus constant on the internal space. We now perform a $\SL{5}$ transformation on the fields of the 7-dimensional supergravity, i.e.
\begin{equation}
 \gM'_{\bar{\cA}\bar{\cB}} = {\cal T}_{\bar{\cA}}{}^{\bar{\cC}} {\cal T}_{\bar{\cB}}{}^{\bar{\CD}} \gM_{\bar{\cA}\bar{\cB}} \,,
\end{equation}
and lift $\gM'$ to 11-dimensional supergravity using the frame fields $\cE'_{\cA}{}^{\bar{\cB}}$. Thus the dual background is described by the fields encoded in the generalised metric
\begin{equation}
 \gM_{\cA\cB} = \cE'_{\cA}{}^{\bar{\cA}} \cE'_{\cB}{}^{\bar{\cB}} \gM'_{\bar{\cA}\bar{\cB}} \,.
\end{equation}

Equivalently, the Poisson-Lie U-duality can be viewed as a local $\SL{5}$ transformation generated by
\begin{equation}
 {\cal T}_{\cA}{}^{\cB} \equiv \cE'_{\cA}{}^{\bar{\cA}}\, {\cal T}_{\bar{\cA}}{}^{\bar{\cB}}\, \left( \cE^{-1} \right)_{\bar{\cB}}{}^{\cA} \,.
\end{equation}
This point of view provides the U-duality analogue of the analysis of \cite{Hassler:2017yza,Demulder:2018lmj,Catal-Ozer:2019hxw,Sakatani:2019jgu,Bugden:2019vlj}.
  
\section{The quadratic constraint} \label{s:QC}
If we are given a Drinfeld extension, as defined in section \ref{s:Drinfeld} with structure constants related by \ref{eq:Ttau}, we can use the results of the previous section to immediately construct the frame fields associated to it, and thus to perform a Poisson-Lie U-duality. However, in practice, we typically want to know when a given algebra $\mathfrak{g}$ can be enlarged into a Drinfeld extension, and, in particular, how to define $\lambda_a$ and $\alpha$ and thus the frame fields \eqref{eq:Frame} given $\mathfrak{g}$, so that we can perform a Poisson-Lie U-duality. To answer this question, we must first study the closure conditions of $\eda$, i.e. the quadratic constraints \eqref{eq:QC}, to understand what conditions $\mathfrak{g}$ imposes on the structure constants of $\eda$.

The quadratic constraints \eqref{eq:QC} imply that the irreducible representations of $F_{\cA\cB,\cC}{}^{\CD}$ must satisfy
\begin{equation} \label{eq:QCIrrep}
\begin{split}
F_{\cA\cB,[\cC}{}^{\cE} \tau_{\CD]\cE} &= 0 \,, \\
F_{\cA\cB,(\cC}{}^{\cE} S_{\CD)\cE} &= 0 \,, \\
2 F_{\cA\cB,\cF}{}^{[\cC} Z^{|\cF|\CD],\cE} + F_{\cA\cB,\cF}{}^{\cE} Z^{\cC\CD,\cF} - \frac12 \tau_{\cA\cB} Z^{\cC\CD,\cE} &= 0 \,.
\end{split}
\end{equation}
These live in the $\mbf{10} \otimes \left( \mbf{10} \oplus \mbf{15} \oplus \obf{40} \right)$. We now use \eqref{eq:EmbgAlg} and \eqref{eq:Ttau} to simplify the quadratic constraints and find:
\begin{itemize}
	\item The conventional Jacobi identity for $\frak{g}$: $f_{[ab}{}^e f_{c]e}{}^d = 0$, as well as
	\begin{equation}
	 0 = f_{ab}{}^c \tau_{c5}\,.
	\end{equation}
	\item The compatibility conditions involving the $\frak{g}$ action on the remaining embedding tensor components, $S_{ab}$ and $\tau_{ab}$:
	\begin{equation}
	\begin{split} \label{eq:CocIrrep} 
		0 &= f_{ab}{}^d S_{cd} - 2 f_{c[a}{}^d S_{b]d} - \frac43 K_{[a} S_{b]c} - \frac43 K_c \tau_{ab}  \,, \\
	0 &= \frac12 I_{[a} \tau_{bc]} + \tau_{[ab} \tau_{c]5} \,,   \\
	0 &= f_{c[a}{}^d \tau_{b]d} + \frac13 K_c \tau_{ab} - \frac12 \tau_{c[a} \tau_{b]5} + \frac14 S_{c[a} \tau_{b]5} \,, \\
	0 &= f_{[ab}{}^{d} S_{c]e} + 2 f_{[ab}{}^{d}\tau_{c]e} + \frac23 S_{e[a} L_{b} \delta_{c]}^d + \frac23 L_e \tau_{[ab} \delta_{c]}^d + 3 \delta_{[c}^d \delta_{ab]e}^{fgh} \tau_{fg} I_{h} \,,
	\end{split} 
	\end{equation} 
	in which we recall $K_a = I_a + \frac12 \tau_{a5}$ and define $L_a = \tau_{a5} - I_a$.
	\item A dual Jacobi condition, involving only $S_{ab}$ and $\tau_{ab}$ and not the structure constants of $\frak{g}$:
	\begin{equation}
	\begin{split} \label{eq:DJacobi}
	0 &= \tau_{[ab} \tau_{cd]} \,, \\
	0 &= S_{a[b} \tau_{cd]} \,.
	\end{split}
	\end{equation}
\end{itemize}

\subsection{Dual structure constants}
It is worthwhile revisiting the above conditions in terms of the dual structure constants \eqref{eq:DualStrConst0}
\begin{equation}
\tilde{f}^{abc}{}_d = -4\, \epsilon^{abce} \left( S_{de} - 2 \tau_{de} \right) \,.
\end{equation}
Now the compatibility conditions \eqref{eq:CocIrrep} are equivalent to
\begin{equation} \label{eq:Cocycle}
 \begin{split}
 6 f_{f[a}{}^{[c} \tilde{f}^{de]f}{}_{b]} + f_{ab}{}^{f} \tilde{f}^{cde}{}_f + \frac23 \tilde{f}^{cde}{}_{[a} L_{b]}  &= 0 \,, \\
 \tilde{f}^{abc}{}_c I_b &= 0 \,, \\
 \tilde{f}^{abc}{}_c \tau_{b5} &= 0 \,, \\
 f_{de}{}^{a} \tilde{f}^{bde}{}_c + \frac23 \tilde{f}^{abd}{}_c L_d &= 0 \,.
  \end{split}
\end{equation}

Finally, the dual Jacobi conditions \eqref{eq:DJacobi} are equivalent to the fundamental identity for $\tilde{f}^{abc}{}_d$:
\begin{equation}\label{eq:coJac}
\tilde{f}^{abg}{}_c\, \tilde{f}^{def}{}_g - 3 \tilde{f}^{g[de}{}_c\, \tilde{f}^{f]ab}{}_g = 0 \,.
\end{equation}
This implies that the dual structure constants $\tilde{f}^{abc}{}_d$ define a Nambu bracket.

\subsection{Cocycle condition}
The first equation in \eqref{eq:Cocycle}
\begin{equation} \label{eq:Coc}
 6 f_{f[a}{}^{[c} \tilde{f}^{de]f}{}_{b]} + f_{ab}{}^{f} \tilde{f}^{cde}{}_f + \frac23 \tilde{f}^{cde}{}_{[a} L_{b]}  = 0 \,,
\end{equation}
is particularly interesting. Note that the dual structure constants define a map: $\tilde{f}: \mathfrak{g} \longrightarrow \Lambda^3 \mathfrak{g}$, given by
\begin{equation}
\tilde{f}(x) = x^a\, \tilde{f}^{bcd}{}_a\, t_b \wedge t_c \wedge t_d \,, \; \forall \, x = x^a\, t_a \in \mathfrak{g} \,.
\end{equation}
Viewed this way $\tilde{f}$ defines a $\Lambda^3 \mathfrak{g}^*$-valued Lie algebra 1-cochain, which is a useful perspective for what follows.

Let us now first focus on the case $L_a = \tau_{a5} - I_a = 0$. Then \eqref{eq:Coc} implies that $\tilde{f}^{abc}{}_d$ must be a Lie-algebra 1-cocycle, i.e. $\tilde{f}$ must be closed under the Lie algebra differential
\begin{equation}
 d\tilde{f}(x, y) \equiv ad_x \tilde{f}(y) - ad_y \tilde{f}(x) - \tilde{f}([x,y]) = 0 \,, \; \forall \, x,\,y \in \mathfrak{g} \,.
\end{equation}
The usual Lie algebra differential is nilpotent $d^2 = 0$, and thus the cocycle condition is solved by a coboundary
\begin{equation}
 \tilde{f} = dm \,,
\end{equation}
for some $m \in \Lambda^3 \mathfrak{g}^\star$, where
\begin{equation}
 dm(x) \equiv ad_x m \,,\; \forall\, x \in \mathfrak{g} \,.
\end{equation}

Now let us turn to $L_a \neq 0$. Now, the modified cocycle condition \eqref{eq:Coc} can be expressed in terms of the operator
\begin{equation}
\begin{split} \label{eq:ModCocycle}
d'\tilde{f}(x,y) &\equiv ad_x \tilde{f}(y) - ad_y \tilde{f}(x) - \tilde{f}([x,y]) - \frac13 \left( \tilde{f}(x) L(y) - \tilde{f}(y) L(x) \right) = 0 \; \,, \forall\, x,\,y \in \mathfrak{g} \,,
\end{split}
\end{equation}
where
\begin{equation}
 L(x) \equiv L_a x^a \,, \; \forall\, x = x^a\, t_a \in \mathfrak{g} \,.
\end{equation}
However, the operator $d'$ is still nilpotent. Let
\begin{equation} \label{eq:dp}
 d'm(x) \equiv ad_x m + \frac13 L(x) m \,, \; \forall\, x \in \mathfrak{g} \,,
\end{equation}
for some $m \in \Lambda^3 \mathfrak{g}^\star$. Then it is easy to show that $d'^2 = 0$. Therefore, the modified cocycle condition \eqref{eq:Coc} can be solved by a coboundary
\begin{equation} \label{eq:Coba}
 \tilde{f} = d'm \,,
\end{equation}
for any $m \in \Lambda^3 \mathfrak{g}^\star$.

\section{Coboundary Ansatz} \label{s:Coboundary}
As discussed above, the quadratic constraints lead to compatibility conditions between the structure constants of $\mathfrak{g}$ and $\tilde{f}^{abc}{}_d$ \eqref{eq:Cocycle}, which include the modified cocycle condition \eqref{eq:Coc}. Let us now focus on coboundary solutions to these equations, i.e. solutions of the form \eqref{eq:Coba}. In the case of Poisson-Lie T-duality, the analogous coboundary case leads to Yang-Baxter deformations. Therefore, when $\tilde{f}$ is a modified coboundary as above this should lead to the U-duality analogue of Yang-Baxter deformations. Moreover, Whitehead's lemma states that for semisimple Lie algebras $H^1(\mathfrak{g}) = 0$, i.e. for semisimple Lie algebras every 1-cocycle is a coboundary, and thus the coboundary Ansatz is the only solution for semisimple Lie algebras with $I_a = \tau_{a5}$.

Writing \eqref{eq:Coba} out explicitly, it becomes
\begin{equation} \label{eq:CoBM}
 \tilde{f}^{abc}{}_d = 12\, f_{de}{}^{[a} m^{bc]e} + \frac43 m^{abc} L_d \,.
\end{equation}
In terms of the irreps $\tau_{ab}$ and $S_{ab}$, the Ansatz \eqref{eq:CoBM} is
\begin{equation} \label{eq:StCoB}
 S_{ab} = \frac83 m_{(a} K_{b)} \,, \qquad \tau_{ab} = \frac43 m_{[a} K_{b]} + 2\, f_{ab}{}^c m_c \,.
\end{equation}

Now the first equation of \eqref{eq:Cocycle} is automatically solved and the remaining compatibility conditions impose the following restrictions on $m_a$:
\begin{equation}
 \begin{split}
   0 &= f_{ab}{}^d f_{cd}{}^e m_e - f_{ab}{}^d K_{c} m_d + f_{c[a}{}^d \tau_{b]5} m_d - \frac49 K_c m_{[a} K_{b]}  - \frac23 K_{c} m_{[a} \tau_{b]5}  \,, \\
   0 &= 2 f_{[ab}{}^d f_{c]e}{}^f m_f + \frac43 f_{[ab}{}^d m_{c]} K_e + \frac49 m_e K_{[a} L_{b} \delta_{c]}^d + \frac49 K_e m_{[a} L_{b} \delta_{c]}^d + \frac49 L_e m_{[a} K_b \delta_{c]}^d \\
   & \quad + \frac23 L_e f_{[a}{}^f \delta_{c]}^d m_f + 2 \delta_{[c}^d \delta_{ab]e}^{fgh} m_f K_g I_h + 3 \delta_{[c}^d \delta_{ab]e}^{fgh} f_{fg}{}^i m_i I_h \,, \\
   0 &= f_{[ab}{}^{d} K_{c]} m_d \,. 
 \end{split}
\end{equation}
Finally, the dual Jacobi conditions imply
\begin{equation}
 \begin{split}
  f_{[ab}{}^e f_{cd]}{}^f m_e\, m_f &= 0 \,, \\
  K_a f_{[bc}{}^e m_{d]}\, m_e &= 0 \,.
 \end{split}
\end{equation}

Now let us construct the object $\lambda_a$ satisfying the first-order constraints \eqref{eq:allfluxes} with $S_{ab}$ and $\tau_{ab}$ given as in \eqref{eq:StCoB}, as well as the group composition properties \eqref{eq:lambda0} and \eqref{eq:groupcomp}. It is given by
\begin{equation} \label{eq:Coboundary}
 (\lambda_g)_a= m_a - (\Xi_g)^\frac{2}{3}  |ad_g|^{-\frac{2}{3}}  (ad_g)_a{}^b m_b \,,
\end{equation}
which clearly satisfies \eqref{eq:lambda0} and \eqref{eq:groupcomp}. Moreover, we can readily calculate the Lie derivative
\begin{equation}
 L_{v_b} \lambda_a \equiv \lambda_{a,b} =  \frac{2}{3} K_b (\lambda - m)_a + f_{ab}{}^c (\lambda - m)_c\,,
\end{equation}
and find that at the identity where $\lambda_a = 0$
\begin{equation}
 S_{ab} = \frac{8}{3} m_{(a}  K_{b)}  \, , \quad  \tau_{ab} = \frac{4}{3} m_{[a} K_{b]} + 2 f_{ab}{}^cm_c\,,
\end{equation}
as required by \eqref{eq:StCoB}.

Now we note that \eqref{eq:allfluxes} is satisfied precisely when\eqref{eq:constancyconditions} holds. However, as shown in section \ref{s:Drinfeld}, these equations follow from adjoint-invariance of the structure constants, and thus from integrating the quadratic constraints \eqref{eq:QC} over $G$. Since we have already satisfied the quadratic constraints \eqref{eq:QC} and ensured that $\lambda_a$ transforms the right way on $G$, our Ansatz \eqref{eq:Coboundary} satisfies \eqref{eq:constancyconditions} and thus \eqref{eq:allfluxes}. 

Finally, for a given group $G$, we can integrate $d\Xi = \frac12 l^a\, \tau_{a5}\, \Xi$ to find the function $\Xi$ appearing via $\alpha$, see equation \eqref{eq:alpha}, in the frame fields \eqref{eq:Frame}.
 
Given the analogy with Poisson-Lie T-duality, it is tempting to speculate that the matrix $m^{abc}$ defined above is related to an M-theoretic analogue of Yang-Baxter deformations. In particular, one may want to view $m: \mathfrak{g} \longrightarrow \Lambda^2 \mathfrak{g}^\star$ and defining an associated 3-bracket. The appropriate notion of closure of the 3-bracket should then be the fundamental identity \eqref{eq:coJac}. We can then follow \cite{Lust:2018jsx} to find the analogue of the classical Yang-Baxter equation. We leave a detailed study of this question for future work.

\section{Some examples} \label{s:Examples}

Let us now consider some examples of the co-boundary case. 

The most obvious example here is $\frak{g} =   \frak{su}(2) +\frak{u}(1)$ in which we choose $T_4$ be the $\frak{u}(1)$ generator and $T_{i}= \tau_i$ be the Pauli matrices for $i=1\dots 3$.  This however is rather trivial, immediately from $f_{ab}{}^c \tau_{c5}= 0 $ we have that $K_i= 0 $ for $i=1,2,3$.  Then from the constraint $f_{[ab}{}^d K_{c]} m_d = 0$ one concludes that $m_i = 0 $ for $i=1,2,3$. From the constancy condition one is left with $0=f_{[a b }{}^d \lambda_{c]}$ but picking say $a=1, b=2 , c=4, d=3$ one finds that the only remaining component $m_4$ is forced to be zero. Of course, since there is a $U(1)$ factor here there can still be solutions for $\lambda$  not of co-boundary type.

Let us consider $\frak{g} = (II) + \frak{u}(1)$  where $(II)$ is the Bianchi$_{II}$ or Heisenberg-Weyl algebra for which the only non-vanishing commutator is
\begin{equation}
 [T_2, T_3] = i  T_1 \,.
\end{equation}
We can choose a parameterisation of $(II) $ as 
\begin{equation}
T_1 = i \begin{pmatrix} 0 & 0 & 1 \\  0 & 0 & 0 \\  0 & 0 & 0 \end{pmatrix} \,, \quad T_2 = i \begin{pmatrix} 0 & 1 & 0 \\  0 & 0 & 0 \\  0 & 0 & 0 \end{pmatrix}  \,, \quad    T_3 = i \begin{pmatrix} 0 & 0 & 0 \\  0 & 0 & 1 \\  0 & 0 & 0 \end{pmatrix}  \,,
\end{equation}
and a corresponding group element (in which the overall phase  accounts for the $U(1)$ factor) 
\begin{equation}
g =  e^{i \frac{\theta}{3} } \begin{pmatrix} 1 & x & y \\  0 & 1 & z \\  0 & 0 & 1 \end{pmatrix} \,. 
\end{equation}

\begin{table}[h!!] 
\label{eq:solsHiesneberg}  
\begin{equation}
\begin{array}{c||c|c|c|c|c|c}
 ~ &  K_a  &  m_a  & \tau_{a5} & \tau_{ab} & S_{ab} & \tilde{f}   
\\  \hline \hline 
 (1) & \{0,0,0,0\} & \left\{m_1,0,0,0\right\} & - & \times & - & \times \\
 (2) & \left\{0,k_2,k_3,0\right\} & \left\{0,m_2,\frac{k_3 m_2}{k_2},0\right\} & \times & - & \times & \times \\
 (3) & \left\{0,k_2,k_3,k_4\right\} & \{0,0,0,0\} & \times & - & - & - \\
 (4) & \{0,0,0,0\} & \left\{0,m_2,m_3,m_4\right\} & - & - & - & - \\
\end{array}
 \end{equation}
 \caption{Solutions for $\frak{g} = (II) + \frak{u}(1)$ in which $\times$ indicates a non-zero embedding tensor component.  Of these note that  $\lambda_a$ is non-zero only for the first two rows.}
 \end{table} 
 
The function $\Xi_g = e^{\frac{1}{3} \theta k_4 - x k_2 -z k_3}  $ which enters into the co-boundary ansatz for $\lambda$ is compatible with the group multiplication law and the differential constraint $d\Xi_g = \frac{1}{2} l^c \tau_{a5} \Xi_g$. 

The one forms and dual vectors are given as 
\begin{equation}
\begin{split}
l^1 = - dy + x d z  \, , \quad l^2 = -dx \, , \quad l^3 =-dz , \, \quad l^4 = \frac{1}{3} d\theta \\
v_1 = - \partial_y \, , \quad v_2 = - \partial_x \, , \quad v_3 = - x\partial_y - \partial_z , \, \quad v_4 = 3 \partial_\theta \, .
\end{split}
\end{equation}
and the adjoint action is 
\begin{equation}
(ad_g)_a{}^b =  \begin{pmatrix} 1  & 0 & 0 &0  \\  -z & 1 & 0 & 0 \\   x & 0 & 1& 0 \\ 0&0&0& 1\end{pmatrix}\, . 
\end{equation}
The various constraint equations discussed above admit a set of solutions, illustrated in table 1, leading to constant embedding tensor components.  Of these $\lambda_a$ is non zero for only the first two entries of the table as can be seen from the above form of the adjoint action.  Explicitly we have the cases
\begin{equation}
\lambda^{(1)} = m_1  \{ 0, z, -x , 0 \} \, ,  \quad \lambda^{(2)} =\frac{m_2}{k_2} \left(1- e^{-\frac{2}{3}\left(x k_2 + z k_3 \right)}  \right)   \{ 0, k_2 , k_3 , 0 \} \, .  
\end{equation}

\section{Discussion and outlook} \label{s:Conclusions}
This work opens a number of interesting lines.  

As indicated in section 4, by a procedure of undressing, $SL(5)$ action and  redressing, one has a map between two generalised metrics giving rise to same lower dimensional supergravity theory and one should like to exploit this as   a solution generating technique.    An exciting opportunity is to explore pragmatic usages of this technique in constructing new holographic supergravity solutions, a programme which becomes richer as the dimensionality of the internal space is increased.

A immediate mechanical task is to scan through the four-dimensional Lie-algebras (see e.g. \cite{Popovych:2003xb}) and classify all the Exceptional Drinfeld Algebras $\eda\subset SL(5) \times \mathbb{R}^+$ thereby providing an exceptional analogue of the classification of six-dimensional Drinfeld doubles \cite{Snobl:2002kq}.   The framework proposed above should admit a ready generalisation to the other exceptional generalised geometries based on $\frak{e}_{n,(n)}$, at least for $n\leq 6$, though the details should be worked out. New features occur at larger $n$: at $n=6$ a new object  $\lambda_{abcdef}$  would enter and modify the structure; at $n=7$ a mixed symmetry object $\lambda^{a_1\dots a_7, a_8}$ must be considered;  at $n=8$ the generalised Lie derivative alone doesn't close  and compensating additional shift symmetries must be incorporated.  Beyond $n=8$ one is dealing with infinite dimensional algebras  which  would be exciting to investigate in this context.     
 
Here we have taken a target space perspective and one might wonder about the implications of this construction on the world-volume of membranes or five-branes in M-theory.     Recall the worldsheet currents ${\cal J}_a$  associated to the $G$ action generated by the vector fields $v_a$  obey a modified conservation law in the case of  Poisson-Lie   NLSMs.  For the case of $SL(5)$ exceptional generalised geometry we need only think of membranes where  a natural expectation is a modified conservation law of the form
\be
d \star {\cal J}_a = \tilde{f}^{bcd}{}_a  {\cal J}_b\wedge  {\cal J}_c \wedge  {\cal J}_d \, . 
\ee
Beyond $SL(5)$, when five-branes should be considered this is likely to be more involved. 

The Poisson-Lie scenario described in section \ref{s:PLReview} can be generalised in two important ways.  First, the assumption of a Drinfeld double can be relaxed to only require a single isotropic subgroup, this set-up captures WZW models as well as their integrable $\lambda$-deformations.     Second, one can take a   reduction of the Poisson-Lie model to  ``dressing cosets''  which allows for   target spaces that are cosets rather than groups (required to define e.g. integrable Yang-Baxter type deformations of $AdS_5 \times S^5$).   One would hope to situate both of these generalisations within the exception generalised geometry setting. 

Here we have seen that how the components of the embedding tensor are realised as encoding structure constants of an exceptional Drinfeld algebra.  One can thus reduce on the  geometry we construct to a lower dimensional gauged supergravity theory.  Here the resultant theories  obtained after Scherk--Schwartz reduction will be maximally  supersymmetric, and so a natural question related to the preceding paragraph, is to obtain the half-maximal or lower supersymmetric analogue construction exploiting the ideas in \cite{Malek:2017njj}.    The converse question is interesting; under what circumstances can a lower dimensional gauged supergravity theory be uplifted to an exceptional Drinfeld algebra?

\section*{Acknowledgements} 
 DCT is supported by a Royal Society University Research Fellowship {\em Generalised Dualities in String Theory and Holography} URF 150185 and in part by STFC grant ST/P00055X/1 and in part by the ``FWO-Vlaanderen'' through the project G006119N and by the Vrije Universiteit Brussel through the Strategic Research Program ``High-Energy Physics''. EM is supported by the ERC Advanced Grant ``Exceptional Quantum Gravity'' (Grant No. 740209). EM would like to thank Swansea University and both authors would like to thank the Mainz Institute for Theoretical Physics (MITP) of the Cluster of Excellence PRISMA+ (Project ID 39083149) and the workshop ``Integrability, duality and beyond'' at Santiago de Compostela for hospitality while part of this work was being completed. DCT wishes to thank Falk Hassler for enlightening discussions on related topics.

\appendix
\section{Conventions}
\label{appendix:Conventions}
 
We consider an algebra $\frak{g}$ (group $G$) with Hermitian generators $t_a$ obeying $[t_a , t_b]= i f_{ab}{}^c t_c$.  We define the trace of the structure constants as $I_a = f_{a b}{}^b$ such that $f_{ab}{}^c =\hat{f}_{ab}{}^c + \frac{2}{3} \delta_{[b}^c I_{a]}$ with $\hat{f}_{ab}{}^b = 0$.   When $\dim \frak{g}= 4$ we can make use of the  identity $\hat{f}_{ef}{}^{[a} \epsilon^{bc] ef}=0$.

We denote by $l^a$ the left-invariant Maurer-Cartan one-forms on $G$, given by $g^{-1} dg = i\, l^a\, t_a$, and by $v_a$ their dual vector fields that generate right translations. They satisfy
\be\label{eq:conventions}
\begin{aligned}
d l^a = \frac{1}{2} f_{bc}{}^a l^b \wedge l^c \ , \quad    \iota_{v_a} l^b = \delta_a{}^b  \ ,  \quad 
 L_{v_a} v_b = - f_{ab}{}^c v_c \ , \quad L_{v_a} l^b = + f_{a c }{}^b l^c  \, , 
  \end{aligned}
\ee
in which $L$ denotes the Lie-derivative that we recall acts on forms according to $L = d \iota + \iota d$.  
We define the adjoint action of $G$ on $\frak{g}$ as
\be
\textrm{ad}_g : t_a  \mapsto g t_a g^{-1} =   (a_{g})_a{}^b t_b \, . 
\ee
Certain manipulations require the identities 
\be\label{eq:da}
d  (ad_{g})_a{}^b = - l^c f_{c a}{}^d  (ad_{g})_d{}^b \ , \quad d |ag_g | = - l^a I_a |ad_g| \ , \quad (ad_{g})_a{}^d (ad_{g})_b{}^e f_{de}{}^c = f_{ab}{}^d (ad_{g})_d{}^c  \, . 
\ee 

\providecommand{\href}[2]{#2}\begingroup\raggedright\endgroup

\end{document}